\documentclass[9pt,shortpaper,twoside,web]{ieeecolor}
\usepackage{generic}
\usepackage{cite}
\usepackage{amsmath,amssymb,amsfonts}
\usepackage{algpseudocode}
\usepackage{graphicx}
\usepackage{textcomp}

\newtheorem{thm}{Theorem}
\newtheorem{rem}[thm]{Remark}
\newtheorem{defn}[thm]{Definition}

\newtheorem{lem}[thm]{Lemma}

\DeclareGraphicsExtensions{.pdf,.png}

\graphicspath{{.}}

\def\BibTeX{{\rm B\kern-.05em{\sc i\kern-.025em b}\kern-.08em
    T\kern-.1667em\lower.7ex\hbox{E}\kern-.125emX}}
\begin{document}
\title{Temporal Parallelization of Bayesian Smoothers}

\author{Simo S\"arkk\"a,~\IEEEmembership{Senior Member,~IEEE,}
        \'Angel F. Garc\'ia-Fern\'andez %
\thanks{S. S\"arkk\"a is with the Department
of Electrical Engineering and Automation, Aalto University, 02150 Espoo, Finland (email: simo.sarkka@aalto.fi).}%
\thanks{\'Angel F. Garc\'ia-Fern\'andez is with the Department of Electrical Engineering and Electronics, University of Liverpool, Liverpool L69 3GJ, United Kingdom (email: angel.garcia-fernandez@liverpool.ac.uk).}}

\maketitle

\begin{abstract}
This paper presents algorithms for temporal parallelization of Bayesian smoothers. We define the elements and the operators to pose these problems as the solutions to all-prefix-sums operations for which efficient parallel scan-algorithms are available. We present the temporal parallelization of the general Bayesian filtering and smoothing equations and specialize them to linear/Gaussian models. The advantage of the proposed algorithms is that they reduce the linear complexity of standard smoothing algorithms with respect to time to logarithmic. 
\end{abstract}


\section{Introduction}

Parallel computing is rapidly transforming from a scientists' computational tool to a general purpose computational paradigm. The availability of affordable massively-parallel graphics processing units (GPUs) as well as widely-available parallel grid and cloud computing systems \cite{Rauber:2013,Cook:2013,Cormen:2009} drive this transformation by bringing parallel computing technology to everyday use. This creates a demand for parallel algorithms that can harness the full power of the parallel computing hardware. %

Stochastic state-space models allow for modeling of time-behaviour and uncertainties of dynamic systems, and they have long been used in various tracking, automation,  communications, and imaging applications \cite{Jazwinski:1970,Bar-Shalom+Li+Kirubarajan:2001,Sarkka:2013,Cappe+Moulines:2005,Kaipio+Somersalo:2005}. More recently, they have also been used as representations of prior information in machine learning setting (see, e.g., \cite{Sarkka+Alvarez+Lawrence:2019}). In all of these applications, the main problem can be mathematically formulated as a state-estimation problem on the stochastic model, where we estimate the unknown phenomenon from a set of noisy measurement data. Given the mathematical problem, the remaining task is to design efficient computational methods for solving the inference problems on large data sets such that they utilize the computational hardware as effectively as possible. 

Bayesian filtering and smoothing methods \cite{Sarkka:2013} provide the classical  \cite{Ho+Lee:1964} solutions to state-estimation problems which are computationally optimal in the sense that their computational complexities are linear with respect to the number of data points. Although these solutions are optimal for single central processing unit (CPU) systems, due to the sequential nature of the algorithms, their complexity remains linear also in parallel multi-CPU systems. However, genuine parallel algorithms are often capable to perform operations in logarithmic number of steps by massive parallelization of the operations. More precisely, their span-complexity \cite{Cormen:2009}, that is, the number of computational steps as measured by a wall-clock, is often logarithmic with respect to the number of data points. However, their total number of operations, the work-complexity, is still linear as all the data points need to be processed.

Despite the long history of state-estimation methods, the existing parallelization methods have concentrated on parallelizing the subproblems arising in Bayesian filtering and smoothing methods, but there is a lack of algorithms that are specifically designed for solving state-estimation problems in parallel architectures. There are, however, some existing approaches for parallelizing Kalman type of filters as well as particle filters. One approach was studied in \cite{Barfoot:2014} and \cite{Grigorievskiy:2017} is to parallelize the corresponding batch formulation, which leads to sub-linear computational methods, because the matrix computations can be parallelized. If the state-space of the Kalman filter is large, it is then possible to speed up the matrix computations via parallelization \cite{Lyster:1997,Evensen:2003}. Particle filters can also be parallelized over the particles \cite{Lee:2010,Rosen:2013} the bottleneck being the resampling step. In some specific cases such as in multiple target tracking \cite{Liggins:1997} it is possible to develop parallelized algorithms by using the structure of the specific problem.

The contribution of this article is to propose a novel general algorithmic framework for parallel computation of Bayesian smoothing solutions for state space models. We also present algorithms for parallelizing the Bayesian filtering solutions, but our focus is in smoothing, because the parallel computation is done off-line in the sense that all the data needs to be available during the parallel computations and it cannot arrive sequentially. Our approach to parallelization differs from the aforementioned approaches in the aspect that we replace the whole Bayesian filtering and smoothing formalism with another, parallelizable formalism. We replace the Bayesian filtering and smoothing equations \cite{Jazwinski:1970,Sarkka:2013} with another set of equations that can be combined with so-called scan or prefix-sums algorithm \cite{Ladner:1980,Blelloch:1989,Blelloch:1990,Cook:2013}, which is one of the fundamental algorithm frameworks in parallel computing. The advantage of this is that it allows for reduction of the linear $O(n)$ complexity of batch filtering and smoothing algorithms to logarithmic $O(\log n)$ span-complexity in the number $n$ of data points. Based on the novel formulation we develop parallel algorithms for computing the filtering and smoothing solutions to linear Gaussian systems with the logarithmic span-complexity. As this parallelization is done in temporal direction, the individual steps of the resulting algorithm could further be parallelized in the same way as Kalman filters and particle filters have previously been parallelized \cite{Lyster:1997,Liggins:1997,Evensen:2003,Lee:2010,Rosen:2013}.

The organization of the article is the following. In Section~\ref{sec:background} we review the classical Bayesian filtering and smoothing methodology as well as the parallel scan algorithm for computing prefix sums. In Section~\ref{sec:Parallel-Bayesian-filtering} we present the general framework for parallelizing Bayesian filtering and smoothing methods. Section~\ref{sec:Parallel-linear-Gaussian-filter} is concerned with specializing the general framework to linear Gaussian systems. Numerical example of linear/Gaussian systems is given in Section~\ref{sec:numerical} and finally Section~\ref{sec:conclu} concludes the article along with discussion on various aspects of the methodology.

\section{Background} \label{sec:background}

\subsection{Bayesian filtering and smoothing} \label{sec:bfs}

Bayesian filtering and smoothing methods \cite{Sarkka:2013} are algorithms for  statistical inference in probabilistic state-space models of the form
\begin{equation}
\begin{split}
  x_{k} &\sim p(x_{k} \mid x_{k-1}), \\
  y_{k} &\sim p(y_{k} \mid x_{k}).
  \label{eq:ssmodel}
\end{split}
\end{equation}
with $x_{0} \sim p(x_{0})$. Above, the state  $x_k\in \mathbb{R}^{n_{x}}$ at time step $k$ evolves as a Markov process with transition density $p(x_{k} \mid x_{k-1})$. State $x_k$ is observed by the measurement $y_k\in \mathbb{R}^{n_{y}}$ whose density is  $p(y_{k} \mid x_{k})$. 

The objective of Bayesian filtering is to compute the posterior density $p(x_k \mid y_{1:k})$ of the state $x_k$ given the measurements  $ y_{1:k}=(y_1,\ldots,y_k)$ up to time step $k$. Given the measurements up to a time step $n$, the objective is smoothing is to compute the density $p(x_k \mid y_{1:n})$ of the state $x_k$ for $k<n$.

The key insight of Bayesian filters and smoothers is that the computation of the required densities can be done in linear $O(n)$ number of computational steps by using recursive (forward) filtering and (backward) smoothing algorithms. This is significant, because a naive computation of the posterior distribution would typically take at least $O(n^3)$ computational steps. 

The Bayesian filter is a sequential algorithm, which iterates the following prediction and update steps:
\begin{equation}
    p(x_k \mid y_{1:k-1})
    = \int p(x_k \mid x_{k-1})
       \, p(x_{k-1} \mid y_{1:k-1}) \, \mathrm{d}x_{k-1},
\label{eq:bayesf_p}
\end{equation}
\begin{equation}
    p(x_k \mid y_{1:k})
    = \frac{p(y_k \mid x_k)
       \, p(x_k \mid y_{1:k-1})}
       {\int p(y_k \mid x_k)
          \, p(x_k \mid y_{1:k-1}) \, \mathrm{d}x_k}.
\label{eq:bayesf_u}
\end{equation}
Given the filtering outputs for  $k=1,\ldots,n$, the Bayesian forward-backward smoother consists of the following backward iteration for $k=n-1,\ldots,1$:
\begin{equation}
\begin{split}
  &p(x_k \mid y_{1:n}) \\
  &= p(x_k \mid y_{1:k})
     \int \frac{p(x_{k+1} \mid x_k)
     	\, p(x_{k+1} \mid y_{1:n})}
     {p(x_{k+1} \mid y_{1:k})}
         \mathrm{d} x_{k+1}.
\end{split}
\label{eq:bsmoothingeqs}
\end{equation}
When applied to a batch of data of size $n$, the computational complexity of the filter and smoother is $O(n)$ as they perform $n$ sequential steps in forward and backward directions when looping over the data. The Kalman filter and Rauch--Tung--Striebel (RTS) smoother \cite{Kalman:1960,Rauch+Tung+Striebel:1965} are the solutions to these recursions when the transition densities are linear and Gaussian. The filtering and smoothing equations can also be analogously solved in closed form for discrete-state models \cite{Cappe+Moulines:2005}. In this paper, we show how to parallelize the previous recursions using parallel scan-algorithm, which is reviewed next.

\subsection{Parallel scan-algorithm} \label{sec:scan}
The parallel-scan algorithm \cite{Blelloch:1990} is a general parallel computing framework that can be used to convert sequential $O(n)$ algorithms with certain associative property to $O(\log n)$ parallel algorithms. The algorithm was originally developed for computing prefix sums \cite{Ladner:1980}, where it uses the associative property of summation. The algorithm has since been generalized to arbitrary associative binary operators and it is used as the basis of multitude of parallel algorithms including sorting, linear programming, and graph algorithms. This kind of algorithms are especially useful in GPU-based computing systems and they are likely to be fundamental algorithms in a many future parallel computing systems.

The problem that the parallel-scan algorithm  \cite{Blelloch:1990} solves is the all-prefix-sums operation, which is defined next.  

\begin{defn} Given a sequence of elements $(a_1, a_2, \ldots, a_n)$, where $a_i$ belongs to a certain set, along with an associative binary operator $\otimes$ on this set, the all-prefix-sums operation computes the sequence 
	\begin{equation}
	(a_1, a_1 \otimes a_2, \ldots, a_1 \otimes \cdots \otimes a_n). 
	\label{eq:allprefixsums}
	\end{equation} 
\end{defn}

For example, if we have $n=4$, $a_i=i$, and $\otimes$ denotes the ordinary summation, the all-prefix-sums are $(1,3,6,10)$. If $\otimes$ denotes the subtraction, the all-prefix-sums are $(1,-1,-4,-8)$. It should be noted that the operator is not necessarily commutative so we use a product symbol, as matrix products are not commutative, instead of a summation symbol.

The all prefix-sums operation can be computed sequentially by processing one element after the other. However, this direct sequential iteration inherently takes $O(n)$ time. We can now see the analogy of the iteration to the Bayesian filter discussed in previous section -- both of the algorithms have linear $O(n)$ complexity, because they need to loop over all the elements in forward direction. A similar argument applies to the Bayesian smoothing pass. 

Fortunately, the all-prefix-sum sums operation can be computed in parallel in $O(\log n)$ span-time by using {\em up-sweep} and {\em down-sweep} algorithms \cite{Blelloch:1990} shown in Fig.~\ref{alg:parprefix}. These algorithms correspond to up and down traversals in a binary tree which are used for computing partial (generalized) sums of the elements. A final pass is then used to construct the final result. The algorithms can be used for computing all-prefix-sums \eqref{eq:allprefixsums} for an arbitrary associative operator $\otimes$. %

\begin{figure}
\begin{algorithmic}
 \State // Save the input:
  \For{$i\gets1$ \textbf{to} $n$} \Comment{Compute in parallel}
      \State $b_i \gets a_i$
  \EndFor
  \State // Up sweep:
  \For{$d\gets0$ \textbf{to} $\log_2 n - 1$} 
    \For{$i\gets0$ \textbf{to} $n-1$ \textbf{by} $2^{d+1}$} \Comment{Compute in parallel}
      \State $j \gets i + 2^d$
      \State $k \gets i + 2^{d+1}$
      \State $a_k \gets a_j \otimes a_k$
    \EndFor
  \EndFor
  \State // Down sweep:
  \For{$d\gets\log_2 n - 1$ \textbf{to} $0$}
    \For{$i\gets0$ \textbf{to} $n-1$ \textbf{by} $2^{d+1}$} \Comment{Compute in parallel}
      \State $j \gets i + 2^d$
      \State $k \gets i + 2^{d+1}$
      \State $t \gets a_j$
      \State $a_j \gets a_k$
      \State $a_k \gets a_k \otimes t$
    \EndFor
  \EndFor
  \State // Final pass:
  \For{$i\gets1$ \textbf{to} $n$} \Comment{Compute in parallel}
       \State $a_i \gets a_i \otimes b_i$
  \EndFor
\end{algorithmic}
\caption{Parallel scan algorithm for in-place transformation of the sequence $(a_i)$ into its all-prefix-sums in $O(\log n)$ span-complexity. Note that the algorithm in this forms assumes that $n$ is a power of $2$, but it can easily be generalized to an arbitrary $n$.}
\label{alg:parprefix}
\end{figure}

\section{Parallel Bayesian filtering and smoothing\label{sec:Parallel-Bayesian-filtering}}

In this section, we explain how to define the elements and the binary
operators to be able to perform Bayesian filtering and smoothing using
parallel scan algorithms. 

\subsection{Bayesian filtering}\label{subsec:Bayesian_filtering}

In order to perform parallel Bayesian filtering, we need to find the
suitable element $a_{k}$ and the binary associative operator $\otimes$.
As we will see in this section, an element $a$ consists of a pair
$\left(f,g\right)\in\mathcal{F}$ where $\mathcal{F}$ is
\begin{equation}
\mathcal{F}=\left\{ \left(f,g\right):\int f\left(y\mid z\right)\mathrm{d}y=1\right\}, \label{eq:set_filtering}
\end{equation}
and $f:\mathbb{R}^{n_{x}}\times\mathbb{R}^{n_{x}}\rightarrow \left[0,\infty\right)$ represents a conditional density, and $g:\mathbb{R}^{n_{x}}\rightarrow \left[0,\infty\right)$ represents a likelihood.
\begin{defn}
	\label{def:Operator_filtering}Given two elements $\left(f_{i},g_{i}\right)\in\mathcal{F}$
	and $\left(f_{j},g_{j}\right)\in\mathcal{F}$, the binary associative
	operator $\otimes$ for Bayesian filtering is
	\begin{align*}
	\left(f_{i},g_{i}\right)\otimes\left(f_{j},g_{j}\right) & =\left(f_{ij},g_{ij}\right),
	\end{align*}
	where
	\begin{align*}
	f_{ij}\left(x\mid z\right) & =\frac{\int g_{j}\left(y\right)f_{j}\left(x\mid y\right)f_{i}\left(y\mid z\right)\mathrm{d}y}{\int g_{j}\left(y\right)f_{i}\left(y\mid z\right)\mathrm{d}y}, \\
	g_{ij}\left(z\right) & =g_{i}\left(z\right)\int g_{j}\left(y\right)f_{i}\left(y\mid z\right)\mathrm{d}y.
	\end{align*}
\end{defn}
The proof that $\otimes$ has the associative property is given in
Appendix~\ref{sec:AppendixA}.
\begin{thm}
	\label{thm:Filtering}Given the element $a_{k}=\left(f_{k},g_{k}\right)\in\mathcal{F}$
	where
	\begin{align*}
	f_{k}\left(x_{k}\mid x_{k-1}\right) & =p\left(x_{k}\mid y_{k},x_{k-1}\right), \\
	g_{k}\left(x_{k-1}\right) & =p\left(y_{k}\mid x_{k-1}\right),
	\end{align*}
	$p\left(x_{1}\mid y_{1},x_{0}\right)=p\left(x_{1}\mid y_{1}\right)$,
	and $p\left(y_{1}\mid x_{0}\right)=p\left(y_{1}\right)$, the $k$-th
	prefix sum is
	\begin{align*}
	a_{1}\otimes a_{2}\otimes\cdots\otimes a_{k} & =\left(\begin{array}{c}
	p\left(x_{k}\mid y_{1:k}\right)\\
	p\left(y_{1:k}\right)
	\end{array}\right).
	\end{align*}
\end{thm}
Theorem~\ref{thm:Filtering}  is proved in Appendix~\ref{sec:AppendixA}.
Theorem~\ref{thm:Filtering} implies that we can parallelise the computation
of all filtering distributions $p\left(x_{k}\mid y_{1:k}\right)$
and the marginal likelihoods $p\left(y_{1:k}\right)$, of which the latter ones can be used
for parameter estimation \cite{Sarkka:2013}. 
\begin{rem}
	If we only know $p\left(y_{k}\mid x_{k-1}\right)$ up to a proportionality
	constant, which means that $g_{k}\left(x_{k-1}\right)\propto p\left(y_{k}\mid x_{k-1}\right)$,
	we can still recover the filtering density $p\left(x_{k}\mid y_{1:k}\right)$
	by the above operations. However, we will not be able to recover
	the marginal likelihoods $p\left(y_{1:k}\right)$. We can nevertheless
	recover $p\left(y_{1:k}\right)$ by an additional parallel pass,
	as will be explained in Section~\ref{subsec:Additional_aspects}.
\end{rem}

\subsection{Bayesian smoothing}

The Bayesian smoothing pass requires that the filtering densities
have been obtained beforehand. In smoothing, we consider a different
type of element $a$ and binary operator $\otimes$ than those used in
filtering. As we will see in this section, an element $a$ is a function
$a:\mathbb{R}^{n_{x}}\times\mathbb{R}^{n_{x}}\rightarrow\left[0,\infty\right)$ that belongs to the set
\[
\mathcal{S}=\left\{ a:\int a\left(x\mid z\right)\mathrm{d}x=1\right\} .
\]
\begin{defn}
	\label{def:Operator_smoothing}Given two elements $a_{i}\in\mathcal{S}$
	and $a_{j}\in\mathcal{S}$, the binary associative operator $\otimes$
	for Bayesian smoothing is
	\begin{align*}
	a_{i}\otimes a_{j} & =a_{ij},
	\end{align*}
	where
	\begin{align*}
	a_{ij}\left(x\mid z\right) & =\int a_{i}\left(x\mid y\right)a_{j}\left(y\mid z\right)\mathrm{d}y.
	\end{align*}
\end{defn}
The proof that $\otimes$ has the associative property is included in Appendix~\ref{sec:AppendixB}. 
\begin{thm}
	\label{thm:Smoothing}Given the element $a_{k}=p\left(x_{k}\mid y_{1:k},x_{k+1}\right)\in\mathcal{S}$,
	with $a_{n}=p\left(x_{n}\mid y_{1:n}\right)$, we have that 
	\begin{align*}
	a_{k}\otimes a_{k+1}\otimes\cdots\otimes a_{n} & =p\left(x_{k}\mid y_{1:n}\right).
	\end{align*}
\end{thm}

Theorem~\ref{thm:Smoothing} is proved in Appendix~\ref{sec:AppendixB}.
Theorem~\ref{thm:Smoothing} implies that we can compute all smoothing
distributions in parallel form. However, it should be noted we should
apply the parallel scan algorithm with elements in reverse other,
that is, with elements $b_{k}=a_{n-k+1}$, so that the prefix-sums
$b_{1}\otimes\cdots\otimes b_{k}$ recover the smoothing densities.

\subsection{Additional aspects}\label{subsec:Additional_aspects}

We proceed to discuss additional aspects of the previous formulation of filtering and smoothing. In Section~\ref{subsec:Bayesian_filtering}, it was indicated
that the marginal likelihood $p\left(y_{1:n}\right)$ is directly
available from the parallel scan algorithm if $g_{k}\left(x_{k-1}\right)=p\left(y_{k}\mid x_{k-1}\right)$.
However, sometimes we only know $p\left(y_{k}\mid x_{k-1}\right)$
up to a proportionality constant so $g_{k}\left(x_{k-1}\right)\propto p\left(y_{k}\mid x_{k-1}\right)$,
as will happen in Section~\ref{sec:Parallel-linear-Gaussian-filter}. Although in this
case, the parallel scan Bayesian filtering algorithm provides us with
the filtering densities but not the marginal likelihood $p\left(y_{1:n}\right)$, we can still recover the marginal likelihoods as follows. We first run the parallel filtering algorithm
to recover all filtering distributions $p\left(x_{k}\mid y_{1:k}\right)$
for $k=1$ to $n$ and then, we perform the following decomposition
for $p\left(y_{1:n}\right)$
\begin{align*}
p\left(y_{1:n}\right) & =\prod_{k=1}^{n}p\left(y_{k}\mid y_{1:k-1}\right),
\end{align*}
where 
\begin{align*}
p\left(y_{k}\mid y_{1:k-1}\right) & = \int p\left(y_{k}\mid x_{k}\right)p\left(x_{k}\mid y_{1:k-1}\right)\mathrm{d}x_{k}.
\end{align*}

Each factor $p\left(y_{k}\mid y_{1:k-1}\right)$ can be computed in parallel using the predictive density $p\left(x_{k}\mid y_{1:k-1}\right)$ and the likelihood $p\left(y_{k}\mid x_{k}\right)$. We can then recover all $p\left(y_{1:k}\right)$ by $O(\log n)$ parallel recursive pairwise multiplications of the adjacent terms. 

It is also possible to perform the parallelization at block level instead of at individual element level. When using the parallel scan algorithm, we do not need to assign each single-measurement element to a single computational node, but instead we can perform initial computations in blocks such that a single node processes a block of measurements before combining the results with other blocks. The results of the blocks can then be used as the elements in the parallel-scan algorithm. This kind of procedure corresponds to selecting the elements for the scan algorithm to consist of blocks of length $l$:
\begin{align*}
a_{k} & =\left(\begin{array}{c}
p\left(x_{lk}\mid y_{l\left(k-1\right)+1:kl},x_{l\left(k-1\right)}\right)\\
p\left(y_{l\left(k-1\right)+1:kl}\mid x_{l\left(k-1\right)}\right)
\end{array}\right)
\end{align*}
in filtering and
\begin{equation}
a_{k}=p\left(x_{lk}\mid y_{1:l\left(k+1\right)-1},x_{l\left(k+1\right)}\right)
\end{equation}
in smoothing instead of the corresponding terms with $l=1$. A practical advantage of this is that we can more easily distribute the computations to a limited number of computational nodes while still getting the optimal speedup from parallelization.

\section{Parallel linear/Gaussian filter and smoother\label{sec:Parallel-linear-Gaussian-filter}}

The parallel linear/Gaussian filter and smoother are obtained by particularising
the element $a$ and binary operator $\otimes$ for Bayesian filtering
and smoothing explained in the previous section to linear/Gaussian
systems. The sequential versions of these algorithms correspond to
the Kalman filter and the RTS smoother. 

We consider the linear/Gaussian state space model 
\begin{align*}
x_{k} & =F_{k-1}x_{k-1}+u_{k-1}+q_{k-1}, \\
y_{k} & =H_{k}x_{k}+d_{k}+r_{k},
\end{align*}
where $F_{k-1}\in\mathbb{R}^{n_{x}\times n_{x}}$ and $H_{k}\in\mathbb{R}^{n_{y}\times n_{x}}$
are known matrices, $u_{k-1}\in\mathbb{R}^{n_{x}}$ and $d_{k}\in\mathbb{R}^{n_{y}}$
are known vectors, and $q_{k-1}$ and $r_{k}$ are zero-mean, independent
Gaussian noises with covariance matrices $Q_{k-1}\in\mathbb{R}^{n_{x}\times n_{x}}$
and $R_{k}\in\mathbb{R}^{n_{y}\times n_{y}}$. The initial distribution is given as $x_0 \sim \mathrm{N}(m_0,P_0)$. With this model, we
have that
\begin{align}
p\left(x_{k}\mid x_{k-1}\right) & =\mathrm{N}\left(x_{k};F_{k-1}x_{k-1}+u_{k-1},Q_{k-1}\right),\label{eq:transition_density_Kalman}\\
p\left(y_{k}\mid x_{k}\right) & =\mathrm{N}\left(y_{k};H_{k}x_{k}+d_{k},R_{k}\right).\label{eq:measurement_density_Kalman}
\end{align}

In this section, we use the notation $\mathrm{N}_{I}\left(\cdot;\eta,J\right)$
to denote a Gaussian density parameterised in information form so
that $\eta$ is the information vector and $J$ is the information
matrix. If a Gaussian distribution has mean $\overline{x}$ and covariance
matrix $P$, its parameterisation in information form is $\eta=P^{-1}\overline{x}$
and $J=P^{-1}$. This parametrization corresponds to so-called information form of Kalman filter \cite{Anderson+Moore:1979}. We also use $I_{n_{x}}$ to denote an identity matrix
of size $n_{x}$.

\subsection{Linear/Gaussian filtering}

We first describe the representation of an element $a_{k}\in\mathcal{F}$
for filtering in linear and Gaussian systems by the following lemma.
\begin{lem}
	\label{lem:Element_linear_filtiering}For linear/Gaussian systems,
	the element $a_{k}\in\mathcal{F}$ for filtering becomes
	\begin{align*}
	f_{k}\left(x_{k}\mid x_{k-1}\right) & =p\left(x_{k}\mid y_{k},x_{k-1}\right)=\mathrm{N}\left(x_{k};A_{k}x_{k-1}+b_{k},C_{k}\right), \\
	g_{k}\left(x_{k-1}\right) & =p\left(y_{k}\mid x_{k-1}\right)\propto\mathrm{N}_{I}\left(x_{k-1};\eta{}_{k},J_{k}\right),
	\end{align*}
	where the parameters of the first term are given for $k > 1$ as
\begin{equation}
\begin{split}
	A_{k} & =\left(I_{n_{x}}-K_{k}H_{k}\right)F_{k-1}, \\
	b_{k} & =u_{k-1}+K_{k}\left(y_{k}-H_{k}u_{k-1}-d_{k}\right), \\
	C_{k} & =\left(I_{n_{x}}-K_{k}H_{k}\right)Q_{k-1}, \\
	K_{k} & =Q_{k-1}H_{k}^{\top}S_{k}^{-1}, \\
	S_{k} & =H_{k}Q_{k-1}H_{k}^{\top}+R_{k},\\
\end{split}
\end{equation}
	and for $k = 1$ as
\begin{equation}
\begin{split}
 m^-_1 &= F_{0} m_{0} + u_{0}, \\
 P^-_1 &= F_{0} P_{0} F^\top_{0} + Q_{0}, \\
   S_1 &= H_1 P^-_1 H_1^\top + R_1, \\
   K_1 &= P^-_1 H^\top_1 S_1^{-1}, \\
   A_1 &= 0, \\
   b_1 &= m^-_1 + K_1 [y_1 - H_1 m^-_1 - d_1], \\
   C_1 &= P^-_1 - K_1 S_1 K_1^\top.
\end{split}
\end{equation}
	The parameters of the second term are given as
\begin{equation}
\begin{split}
	\eta_{k} & =F_{k-1}^{\top}H_{k}^{\top}S_{k}^{-1}\left(y_{k}-H_{k}u_{k-1}-d_{k}\right), \\
	J_{k} & =F_{k-1}^{\top}H_{k}^{\top}S_{k}^{-1}H_{k}F_{k-1},
\end{split}
\end{equation}
	for $k=1,\ldots,n$.
\end{lem}
In Lemma~\ref{lem:Element_linear_filtiering}, densities $p\left(x_{k}\mid y_{k},x_{k-1}\right)$
and $p\left(y_{k}\mid x_{k-1}\right)$ are obtained by applying the
Kalman filter update with measurement $y_{k}$, distributed according
to \eqref{eq:measurement_density_Kalman}, applied to the density
$p\left(x_{k}\mid x_{k-1}\right)$ in \eqref{eq:transition_density_Kalman} and matching the terms. For the first step we have applied the Kalman filter prediction and update steps starting from $x_0 \sim \mathcal{N}(m_0,P_0)$ and matched the terms.

Therefore, an element $a_{k}$ can be parameterised by $\left(A_{k},b_{k},C_{k},\eta_{k},J_{k}\right)$,
which can be computed for each element in parallel. Also, it is relevant
to notice that if the system parameters $\left(F_{k},u_{k},Q_{k},H_{k},d_{k},R_{k}\right)$
do not depend on the time step $k$, the only parameters of $a_{k}$
that depend on $k$ are $b_{k}$ and $\eta_{k}$, as they depend on
the measurement $y_{k}$. 
\begin{lem}
	\label{prop:Operator_linear_filtering}Given two elements $\left(f_{i},g_{i}\right)\in\mathcal{F}$
	and $\left(f_{j},g_{j}\right)\in\mathcal{F}$, with parameterisations
	\begin{align*}
	f_{i}\left(y\mid z\right) & =\mathrm{N}\left(y;A_{i}z+b_{i},C_{i}\right),\\
	g_{i}\left(z\right) & \propto\mathrm{N}_{I}\left(z;\eta{}_{i},J_{i}\right),\\
	f_{j}\left(y\mid z\right) & =\mathrm{N}\left(y;A_{j}z+b_{j},C_{j}\right),\\
	g_{j}\left(z\right) & \propto\mathrm{N}_{I}\left(z;\eta{}_{j},J_{j}\right),
	\end{align*}
	the binary operator $\otimes$ for filtering becomes
	\begin{align*}
	\left(f_{i},g_{i}\right)\otimes\left(f_{j},g_{j}\right) & =\left(f_{ij},g_{ij}\right),
	\end{align*}
	where
	\begin{align}
	f_{ij}\left(x\mid z\right) & =\mathrm{N}\left(x;A_{ij}z+b_{ij},C_{ij}\right), \label{eq:lin_f}\\
	g_{ij}\left(z\right) & \propto\mathrm{N}_{I}\left(z;\eta_{ij},J_{ij}\right),
	\label{eq:lin_g}
	\end{align}
	with
	\begin{align*}
	A_{ij} & =A_{j}\left(I_{n_{x}}+C_{i}J_{j}\right)^{-1}A_{i},\\
	b_{ij} & =A_{j}\left(I_{n_{x}}+C_{i}J_{j}\right)^{-1}\left(b_{i}+C_{i}\eta_{j}\right)+b_{j},\\
	C_{ij} & =A_{j}\left(I_{n_{x}}+C_{i}J_{j}\right)^{-1}C_{i}A_{j}^{\top}+C_{j},\\
	\eta_{ij} & =A_{i}^{\top}\left(I_{n_{x}}+J_{j}C_{i}\right)^{-1}\left(\eta_{j}-J_{j}b_{i}\right)+\eta_{i},\\
	J_{ij} & =A_{i}^{\top}\left(I_{n_{x}}+J_{j}C_{i}\right)^{-1}J_{j}A_{i}+J_{i}.
	\end{align*}
\end{lem}
The proof is provided in Appendix~\ref{sec:AppendixC}.

\subsection{Linear/Gaussian smoothing}

We first describe the representation of an element $a_{k}\in\mathcal{S}$
for smoothing in linear and Gaussian systems by the following lemma.
\begin{lem}
	\label{lem:Element_linear_smoothing}For linear/Gaussian systems,
	the element $a_{k}\in\mathcal{S}$ for smoothing becomes
	\begin{align*}
	a_{k}\left(x_{k}\mid x_{k+1}\right) & =p\left(x_{k}\mid y_{1:k},x_{k+1}\right)\\
	& =\mathrm{N}\left(x_{k};E_{k}x_{k+1}+g_{k},L_{k}\right),
	\end{align*}
	where for $k < n$
	\begin{align*}
	E_{k} & =P_{k}F_{k}^{\top}\left(F_{k}P_{k}F_{k}^{\top}+Q_{k}\right)^{-1},\\
	g_{k} & =\overline{x}_{k}-E_{k}\left(F_{k}\overline{x}_{k}+u_{k}\right),\\
	L_{k} & =P_{k}-E_{k}F_{k}P_{k},
	\end{align*}
	and for $k=n$ we have
	\begin{align*}
	E_{n} & = 0, \\
	g_{n} & = \overline{x}_{n},\\
	L_{n} & = P_{n}.
	\end{align*}	
	Above, $\overline{x}_{k}$ and $P_{k}$ are the filtering mean and covariance
	matrix at time step $k$, such that $p\left(x_{k}\mid y_{1:k}\right)=\mathrm{N}\left(x_{k};\overline{x}_{k},P_{k}\right)$. 
\end{lem}
Lemma~\ref{lem:Element_linear_smoothing} is obtained by performing
a Kalman filter update on density $p\left(x_{k}\mid y_{1:k}\right)$
with an observation $x_{k+1}$, whose distribution is given by \eqref{eq:transition_density_Kalman}.
Element $a_{k}$ for smoothing with linear/Gaussian systems can be
parameterised as $a_{k}=\left(E_{k},g_{k},L_{k}\right)$.
\begin{lem}
	\label{prop:Operator_linear_smoothing}Given two elements $a_{i}\in\mathcal{S}$ and $a_{j}\in\mathcal{S}$
	with parameterisation 
	\begin{align*}
	a_{i}\left(y\mid z\right) & =\mathrm{N}\left(y;E_{i}z+g_{i},L_{i}\right),
	\end{align*}
	the binary operator $\otimes$ for smoothing becomes
	\begin{align*}
	a_{i}\otimes a_{j} & =a_{ij},
	\end{align*}
	where
	\begin{align*}
	a_{ij}\left(x\mid z\right) & =\int a_{i}\left(x\mid y\right)a_{j}\left(y\mid z\right)\mathrm{d}y\\
	& =\int\mathrm{N}\left(x;E_{i}y+g_{i},L_{i}\right)\mathrm{N}\left(y;E_{j}z+g_{j},L_{j}\right)\mathrm{d}y\\
	& =\mathrm{N}\left(x;E_{ij}z+g_{ij},L_{ij}\right),
	\end{align*}
	and
	\begin{align*}
	E_{ij} & =E_{i}E_{j},\\
	g_{ij} & =E_{i}g_{j}+g_{i},\\
	L_{ij} & =E_{i}L_{j}E_{i}^{\top}+L_{i}.
	\end{align*}
\end{lem}

\section{Numerical experiment\label{sec:numerical}}

In order to illustrate the benefit of parallelization we consider a simple tracking model (see, e.g., \cite{Bar-Shalom+Li+Kirubarajan:2001,Sarkka:2013}) with the state $x = \begin{pmatrix}  u & v & \dot{u} & \dot{v} \end{pmatrix}^\top$, where $(u,v)$ is the 2D position and $(\dot{u},\dot{v})$ is the 2D velocity of the tracked object. From noisy measurements of the position $(u,v)$, we aim to solve the smoothing problem in order to determine the whole trajectory of the target. 

The model has the form
\begin{equation}
\begin{split}
  x_k &=  F \, x_{k-1} + q_{k-1},\\
  y_k &=  H \, x_k + r_k,
\end{split}
\label{eq:car_model1}
\end{equation}
where $q_k \sim \mathrm{N}(0,Q)$, $r_k \sim \mathrm{N}(0,R)$, and
 \begin{equation}
  \begingroup %
  \setlength\arraycolsep{3pt}
  F = \begin{pmatrix}
    1 & 0 & \Delta t & 0 \\
    0 & 1 & 0 & \Delta t \\
    0 & 0 & 1 & 0 \\
    0 & 0 & 0 & 1
  \end{pmatrix}, \quad
  Q = q \, \begin{pmatrix}
     \frac{\Delta t^3}{3} & 0 & \frac{\Delta t^2}{2} & 0 \\
     0 & \frac{\Delta t^3}{3} & 0 & \frac{\Delta t^2}{2} \\
     \frac{\Delta t^2}{2} & 0 & \Delta t & 0     \\ 
     0 & \frac{\Delta t^2}{2} & 0 & \Delta t
  \end{pmatrix},
  \endgroup
\label{eq:car_model2}
\end{equation}
along with
\begin{equation}
  H = \begin{pmatrix}
    1 & 0 & 0 & 0 \\
    0 & 1 & 0 & 0 
  \end{pmatrix}, \qquad
  R = \begin{pmatrix}
    \sigma^2 & 0  \\
    0 & \sigma^2 
  \end{pmatrix}.
\end{equation}
In our simulations we used the parameters $\sigma = 0.5$, $\Delta t = 0.1$, $q = 1$, and started the trajectory from a random Gaussian initial condition with mean $m_0 = \begin{pmatrix} 0 & 0 & 1 & -1 \end{pmatrix}^\top$ and covariance $P_0 = I_4$.

\begin{figure}
\centerline{\includegraphics{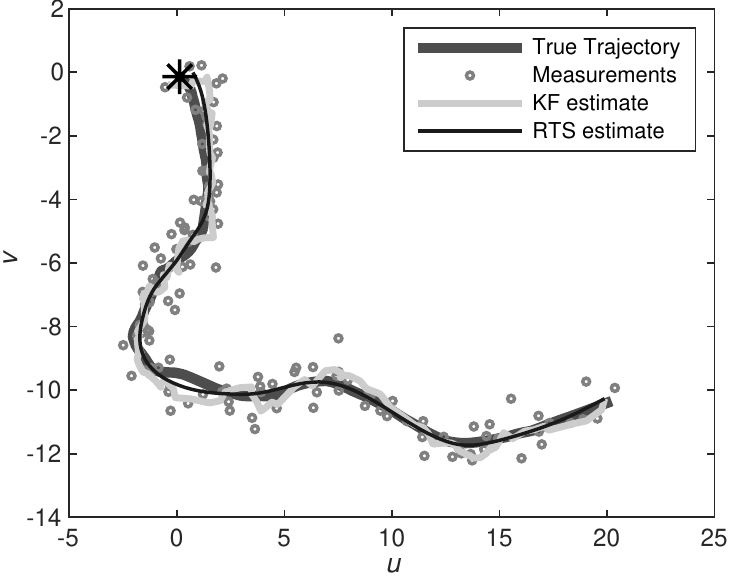}}
\caption{Simulated trajectory from the linear tracking model in Eqs.~\eqref{eq:car_model1} and \eqref{eq:car_model2} along with the Kalman filter (KF) and RTS smoother results.}
\label{fig:car_track}
\end{figure}

Fig.~\ref{fig:car_track} shows a typical trajectory and measurements from the model defined by Eqs.~\eqref{eq:car_model1} and \eqref{eq:car_model2} along with the Kalman filter and RTS smoother solutions. As the parallel algorithms produce exactly the same filter and smoothing solutions as the classic sequential algorithms, this result also illustrates the typical result produced by the proposed algorithms.

\begin{figure}
\centerline{\includegraphics{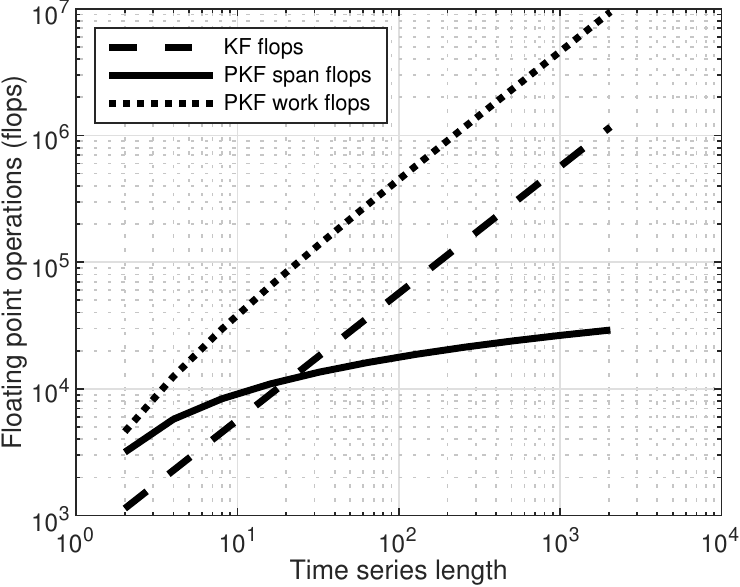}}
\caption{The flops and the span and work flops for the sequential Kalman filter (KF) and the parallel Kalman filter (PKF).}
\label{fig:kf_flops}
\end{figure}

We now aim to evaluate the required number of floating point operations (flops) for generating the smoothing solution for this model. In order to do that, we run the sequential filter and smoothing methods (KF and RTS) as well as the proposed parallel algorithms (PKF and PRTS) over simulated data sets of different sizes and evaluate their span and work flops. The span flops here refers to the minimum number of floating point steps when the parallelizable operations in the algorithm are done in parallel -- this corresponds to the actual execution time required to do the computations in a parallel computer. The work flops refers to the total number of operations that the parallel computer needs to perform -- it measures the total energy required for the computations or equivalently the time required by the algorithm in a single-core computer. As the classic sequential KF and RTS algorithms are not parallelizable, their span and work flops are equal. The flops have been computed by estimating how many flops each of the matrix operations takes (multiplication, summation, LU-factorization) and incrementing the flops counter after every operation in the code. 	

\begin{figure}
\centerline{\includegraphics{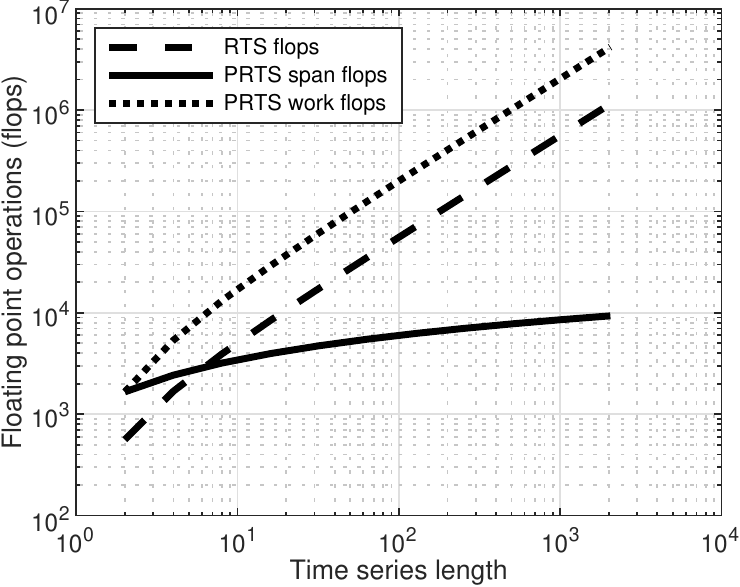}}
\caption{The flops and the span and work flops for the (sequential) RTS smoother and the parallel RTS (PRTS) smoother.}
\label{fig:rts_flops}
\end{figure}

Fig.~\ref{fig:kf_flops} shows the flops required by the sequential KF along with the span flops and work flops required by the parallel Kalman filter algorithm. As expected, with small data set sizes the number of span flops required by the parallel KF is larger than that of the sequential KF, but already starting from time step count of around 20, the span flops is lower for the parallel KF. The logarithmic growth of the span flops in the parallel algorithm can be clearly seen while the number of flops for the sequential KF grows linearly. However, the work flops required by the parallel KF is approximately 8 times the flops of the sequential KF. This means that although the execution time for the parallel algorithms is smaller than for the sequential algorithms, they need to perform more floating point operations in total. 

\begin{figure}
\centerline{\includegraphics{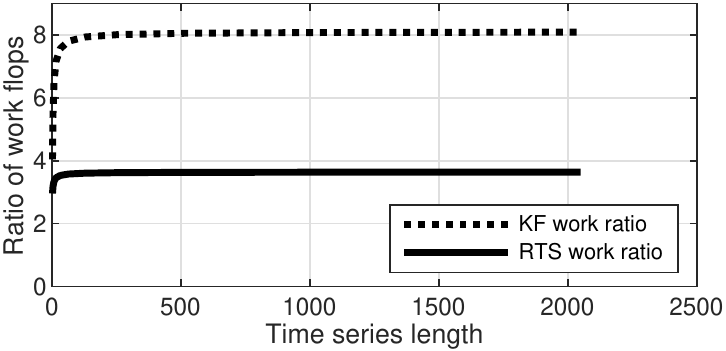}}
\caption{Ratio of work flops for the parallel and sequential Kalman filter and the parallel and sequential RTS smoother.}
\label{fig:kfrts_ratio}
\end{figure}

The flops required by the sequential RTS smoother along with the span flops and work flops required by the parallel RTS smoother are shown in Fig.~\ref{fig:rts_flops}. In this case, the parallel algorithm reaches the sequential algorithm speed already with data set of size less than 10. Furthermore, the total number of floating point operations required by the parallel algorithm is approximately 4 times the operations required by the sequential algorithm. The ratios of these total (work) operations for both the filter and smoother are shown in Fig.~\ref{fig:kfrts_ratio}.

\section{Conclusion and Discussion} \label{sec:conclu}

In this article we have proposed a novel general algorithmic framework for parallel computation of batch Bayesian filtering and smoothing solutions for state space models. The framework is based on formulating the computations in terms of associative operations between suitably defined elements such that the all-prefix-sums operation computed by a parallel-scan algorithm exactly produces the Bayesian filtering and smoothing solutions. The advantage of the framework is that the parallelization allows for performing the computations in $O(\log n)$ span complexity, where $n$ is the number of data points, while sequential filtering and smoothing algorithms have an $O(n)$ complexity. Parallel versions of Kalman filters and Rauch--Tung--Striebel smoothers were derived as special cases of the framework. The computational advantages of the framework were illustrated in a numerical simulation.

A disadvantage of the proposed methodology is that although the wall-clock time of execution is significantly reduced, the total amount of computations (and hence required energy) is larger than with conventional sequential algorithms. Although the total amount of computations is only increased by a constant factor, in some systems, such as small-scale mobile systems, even if parallelization would be possible, it can be beneficial to use the classic algorithms. However, the speedup gain of the proposed approach is beneficial in applications such as data-assimilation based weather forecasting \cite{Cressie::2011} and other spatio-temporal systems appearing, for example, in tomographic reconstruction \cite{Kaipio+Somersalo:2005} or machine learning \cite{Sarkka+Solin+Hartikainen:2013}, where the computations take a significant amount of time. In these systems, it is possible to dedicate the required amount of extra computational resources to gain the significant speedup provided by parallelization. 

Although we have restricted our consideration to specific types parallel-scan algorithms, it is also possible to use other kinds of algorithms for computing the prefix sums corresponding to the Bayesian filtering and smoothing solutions. We could also select algorithms for given computer or network architectures, for minimizing the communication between the nodes, or for minimizing the energy consumption \cite{Grama:2003,Sanders:2006}. The present formulation of the computations in terms of local associative operations is likely to have other applications beyond parallelization. For example, in decentralized systems,  it is advantageous to be able to first perform operations locally and then combine them to produce the full state-estimation solution. 

The proposed framework is also valid for discrete state spaces as well as for other state spaces provided that we consider the elements with the appropriate domain and replace the Lebesgue integrals by integrals with respect to the corresponding reference measure, e.g., counting measure in the case of discrete-state models.

The framework could be extended to non-linear and non-Gaussian models by replacing the exact Kalman filters and smoothers with iterated extended Kalman filters and smoothers \cite{Bell:1993,Bell:1994} or their sigma-point/numerical-integration versions such as posterior linearization filters and smoothers \cite{Garcia:2015,Garcia:2017,Tronarp:2017}. Possible future work also includes developing particle filter and smoother methods (see, e.g., \cite{Sarkka:2013}) for the present framework along with various other Bayesian filter and smoother approximations proposed in literature.

\appendices

\section{\label{sec:AppendixA}}

In this appendix, we prove the required results for Bayesian filtering:
the associative property of the operator in Definition~\ref{def:Operator_filtering}
and Theorem~\ref{thm:Filtering}. 

\subsection{Associative property}

In order to prove the associative property of $\otimes$ for filtering,
we need to prove that for three elements $\left(f_{i},g_{i}\right)$,
$\left(f_{j},g_{j}\right)$, $\left(f_{k},g_{k}\right)\in\mathcal{F}$, the following relation holds
\begin{align}
& \left[\left(f_{i},g_{i}\right)\otimes\left(f_{j},g_{j}\right)\right]\otimes\left(f_{k},g_{k}\right)\nonumber \\
& \:=\left(f_{i},g_{i}\right)\otimes\left[\left(f_{j},g_{j}\right)\otimes\left(f_{k},g_{k}\right)\right].\label{eq:associative_filtering_app}
\end{align}
We proceed to perform the calculations on both sides of the equation
to check that they yield the same result.

\subsubsection{Left-hand side}

We use Definition~\ref{def:Operator_filtering} in the left-hand side
of \eqref{eq:associative_filtering_app} and obtain
\begin{align*}
\left(f_{ij},g_{ij}\right)\otimes\left(f_{k},g_{k}\right) & =\left(f_{ijk},g_{ijk}\right),
\end{align*}
where 
\begin{align}
& f_{ijk}\left(x\mid z\right)\nonumber \\
& =\frac{\iint g_{k}\left(y\right)f_{k}\left(x\mid y\right)g_{j}\left(y'\right)f_{j}\left(y\mid y'\right)f_{i}\left(y'\mid z\right)\mathrm{d}y'\mathrm{d}y}{\iint g_{k}\left(y\right)g_{j}\left(y'\right)f_{j}\left(y\mid y'\right)f_{i}\left(y'\mid z\right)\mathrm{d}y'\mathrm{d}y}, \label{eq:associative_filtering_f_app}
\end{align}
and 
\begin{align}
& g_{ijk}\left(z\right)\nonumber \\
& =g_{ij}\left(z\right)\int g_{k}\left(y\right)f_{ij}\left(y\mid z\right)\mathrm{d}y\nonumber \\
& =g_{i}\left(z\right)\left[\int g_{j}\left(y\right)f_{i}\left(y\mid z\right)\mathrm{d}y\right]\nonumber \\
& \quad\times\left[\int g_{k}\left(y\right)\left[\frac{\int g_{j}\left(y'\right)f_{j}\left(y\mid y'\right)f_{i}\left(y'\mid z\right)\mathrm{d}y'}{\int g_{j}\left(y'\right)f_{i}\left(y'\mid z\right)\mathrm{d}y'}\right]\mathrm{d}y\right]\nonumber \\
& =g_{i}\left(z\right)\iint g_{k}\left(y\right)g_{j}\left(y'\right)f_{j}\left(y\mid y'\right)f_{i}\left(y'\mid z\right)\mathrm{d}y'\mathrm{d}y. \label{eq:associative_filtering_g_app}
\end{align}

\subsubsection{Right-hand side}

We first use operator $\otimes$ to the elements with indices $j$
and $k$ in the right-hand side of \eqref{eq:associative_filtering_app},
see Definition~\ref{def:Operator_filtering}, 
\begin{align*}
\left(f_{j},g_{j}\right)\otimes\left(f_{k},g_{k}\right) & =\left(f_{jk},g_{jk}\right),
\end{align*}
where
\begin{align*}
f_{jk}\left(x\mid z\right) & =\frac{\int g_{k}\left(y\right)f_{k}\left(x\mid y\right)f_{j}\left(y\mid z\right)\mathrm{d}y}{\int g_{k}\left(y\right)f_{j}\left(y\mid z\right)\mathrm{d}y}, \\
g_{jk}\left(z\right) & =g_{j}\left(z\right)\int g_{k}\left(y\right)f_{j}\left(y\mid z\right)\mathrm{d}y.
\end{align*}

Then, the right-hand side of \eqref{eq:associative_filtering_app}
becomes
\begin{align*}
\left(f_{i},g_{i}\right)\otimes\left(f_{jk},g_{jk}\right) & =\left(f'_{ijk},g'_{ijk}\right),
\end{align*}
where
\begin{align}
& f'_{ijk}\left(x\mid z\right)\nonumber \\
& =\frac{\int g_{jk}\left(y\right)f_{jk}\left(x\mid y\right)f_{i}\left(y\mid z\right)\mathrm{d}y}{\int g_{jk}\left(y\right)f_{i}\left(y\mid z\right)\mathrm{d}y}\nonumber \\
& =\frac{\int g_{j}\left(y\right)\left[\int g_{k}\left(y'\right)f_{k}\left(x\mid y'\right)f_{j}\left(y'\mid y\right)\mathrm{d}y'\right]f_{i}\left(y\mid z\right)\mathrm{d}y}{\int\left[g_{j}\left(y\right)\int g_{k}\left(y'\right)f_{j}\left(y'\mid y\right)\mathrm{d}y'\right]f_{i}\left(y\mid z\right)\mathrm{d}y}\nonumber \\
& =\frac{\iint g_{j}\left(y\right)g_{k}\left(y'\right)f_{k}\left(x\mid y'\right)f_{j}\left(y'\mid y\right)f_{i}\left(y\mid z\right)\mathrm{d}y'\mathrm{d}y}{\iint g_{j}\left(y\right)g_{k}\left(y'\right)f_{j}\left(y'\mid y\right)f_{i}\left(y\mid z\right)\mathrm{d}y'\mathrm{d}y},\label{eq:eq:associative_filtering_f_prime_app}
\end{align}
and 
\begin{align}
& g'_{ijk}\left(z\right)\nonumber \\
& =g_{i}\left(z\right)\int g_{jk}\left(y\right)f_{i}\left(y\mid z\right)\mathrm{d}y\nonumber \\
& =g_{i}\left(z\right)\int\left[g_{j}\left(y\right)\int g_{k}\left(y'\right)f_{j}\left(y'\mid y\right)\mathrm{d}y'\right]f_{i}\left(y\mid z\right)\mathrm{d}y\nonumber \\
& =g_{i}\left(z\right)\iint g_{j}\left(y\right)g_{k}\left(y'\right)f_{j}\left(y'\mid y\right)f_{i}\left(y\mid z\right)\mathrm{d}y'\mathrm{d}y.\label{eq:associative_filtering_g_prime_app}
\end{align}
In \eqref{eq:associative_filtering_f_app}, \eqref{eq:associative_filtering_g_app},
it is met that $f'_{ijk}\left(x\mid z\right)=f{}_{ijk}\left(x\mid z\right)$
and $g'_{ijk}\left(z\right)=g{}_{ijk}\left(z\right)$, which proves
the associative property of $\otimes$ in Definition~\ref{def:Operator_filtering}.

\subsection{Proof of Theorem~\ref{thm:Filtering}}

In this appendix, we prove Theorem~\ref{thm:Filtering}. We first
prove by induction that 
\begin{align}
& a_{k-l}\otimes\cdots\otimes a_{k-1}\otimes a_{k}\nonumber \\
& =\left(p\left(x_{k}\mid y_{k-l:k},x_{k-l-1}\right),p\left(y_{k-l:k}\mid x_{k-l-1}\right)\right),\label{eq:theorem_filt_app1}
\end{align}
for $l<k+1$. Relation \eqref{eq:theorem_filt_app1} holds for $l=0$
by definition of $a_{k}$. Then, assuming that 
\begin{align}
& a_{k-l+1}\otimes\cdots\otimes a_{k-1}\otimes a_{k}\nonumber \\
& =\left(p\left(x_{k}\mid y_{k-l+1:k},x_{k-l}\right),p\left(y_{k-l+1:k}\mid x_{k-l}\right)\right)\label{eq:theorem_filt_app2}
\end{align}
holds, we need to prove that \eqref{eq:theorem_filt_app1} holds.

We calculate the first element of $a_{k-l}\otimes b_{k-l+1}$, denoted by
$f_{ab}$, where $b_{k-l+1}=a_{k-l+1}\otimes\cdots\otimes a_{k-1}\otimes a_{k}$.
We have 
\begin{align*}
& f_{ab}\left(x_{k}\mid x_{k-l}\right)\\
& =\frac{\int p\left(y_{k-l+1:k},x_{k}\mid x_{k-l}\right)p\left(x_{k-l}\mid y_{k-l},x_{k-l-1}\right)\mathrm{d}x_{k-l}}{p\left(y_{k-l+1:k}\mid y_{k-l},x_{k-l-1}\right)}\\
& =\frac{p\left(y_{k-l+1:k},x_{k}\mid y_{k-l},x_{k-l-1}\right)}{p\left(y_{k-l+1:k}\mid y_{k-l},x_{k-l-1}\right)}\\
& =p\left(x_{k}\mid y_{k-l:k},x_{k-l-1}\right).
\end{align*}
Function $f_{ab}$ corresponds to the first element of \eqref{eq:theorem_filt_app1},
as required. We further get
\begin{align*}
g_{ab}\left(x_{k-l-1}\right) & =p\left(y_{k-l}\mid x_{k-l-1}\right)\int p\left(y_{k-l+1:k}\mid x_{k-l}\right)\\
& \quad\times p\left(x_{k-l}\mid y_{k-l},x_{k-l-1}\right)\mathrm{d}x_{k-l}\\
& =p\left(y_{k-l}\mid x_{k-l-1}\right)p\left(y_{k-l+1:k}\mid y_{k-l},x_{k-l-1}\right)\\
& =p\left(y_{k-l:k}\mid x_{k-l-1}\right).
\end{align*}
Function $g_{ab}$ corresponds to the second element of \eqref{eq:theorem_filt_app1},
as required.

Substituting $l=k+2$ into \eqref{eq:theorem_filt_app1}, we obtain
\begin{align}
& a_{2}\otimes\cdots\otimes a_{k}\nonumber \\
& =\left(p\left(x_{k}\mid y_{2:k},x_{1}\right),p\left(y_{2:k}\mid x_{1}\right)\right).\label{eq:theorem_filt_app3}
\end{align}
We now calculate the first element of $a_{1}\otimes\left[a_{2}\otimes\cdots\otimes a_{k}\right]$,
denoted as $f_{1k}$, where $a_{1}$ is given in Theorem~\ref{thm:Filtering}:
\begin{align}
& f_{1k}\left(x_{k}\mid x_{0}\right)\nonumber \\
& =\frac{\int p\left(y_{2:k}\mid x_{1}\right)p\left(x_{k}\mid y_{2:k},x_{1}\right)p\left(x_{1}\mid y_{1}\right)\mathrm{d}x_{1}}{\int p\left(y_{2:k}\mid x_{1}\right)p\left(x_{1}\mid y_{1}\right)\mathrm{d}x_{1}}\nonumber \\
& =\frac{\int p\left(y_{2:k},x_{k}\mid x_{1},y_{1}\right)p\left(x_{1}\mid y_{1}\right)\mathrm{d}x_{1}}{p\left(y_{2:k}\mid y_{1}\right)}\nonumber \\
& =\frac{p\left(y_{2:k},x_{k}\mid y_{1}\right)}{p\left(y_{2:k}\mid y_{1}\right)}\nonumber \\
& =p\left(x_{k}\mid y_{1:k}\right).\label{eq:theorem_filt_app4}
\end{align}
The second element of $a_{1}\otimes\left[a_{2}\otimes\cdots\otimes a_{k}\right]$,
denoted as $g_{1k}$ is 
\begin{align}
& g_{1k}\left(x_{0}\right)\nonumber \\
& =p\left(y_{1}\right)\int p\left(y_{2:k}\mid x_{1}\right)p\left(x_{1}\mid y_{1}\right)\mathrm{d}x_{1}\nonumber \\
& =p\left(y_{1:k}\right).\label{eq:theorem_filt_app5}
\end{align}
Results \eqref{eq:theorem_filt_app4} and \eqref{eq:theorem_filt_app5}
finish the proof of Theorem~\ref{thm:Filtering}. 

\section{\label{sec:AppendixB}}

In this appendix, we prove the required results for Bayesian smoothing:
the associative property of the operator in Definition~\ref{def:Operator_smoothing}
and Theorem~\ref{thm:Smoothing}. 

\subsection{Associative property}

In order to prove the associative property of $\otimes$ for filtering,
we need to prove that, for three elements $a_{i}$, $a_{j}$, $a_{k}\in\mathcal{S}$
, the following relation holds:
\begin{align}
\left[a_{i}\otimes a_{j}\right]\otimes a_{k} & =a_{i}\otimes\left[a_{j}\otimes a_{k}\right].\label{eq:associative_smoothing_app}
\end{align}
We proceed to perform the calculations on both sides of the equation
to check that they yield the same result.

\subsubsection{Left-hand side}

We apply the operator in Definition~\ref{def:Operator_smoothing}
on the left-hand side of \eqref{eq:associative_smoothing_app} to
obtain 
\begin{align*}
a_{ij}\otimes a_{k} & =a_{ijk},
\end{align*}
where 
\begin{align}
a_{ijk}\left(x\mid z\right) & =\int a_{ij}\left(x\mid y\right)a_{k}\left(y\mid z\right)\mathrm{d}y\nonumber \\
& =\iint a_{i}\left(x\mid y'\right)a_{j}\left(y'\mid y\right)a_{k}\left(y\mid z\right)\mathrm{d}y\mathrm{d}y'.\label{eq:a_ijk_smoothing_app}
\end{align}

\subsubsection{Right-hand side}

We first calculate $a_{j}\otimes a_{k}$ using the operator in Definition
\ref{def:Operator_smoothing} which gives
\begin{align*}
a_{jk}\left(x\mid z\right) & =\int a_{j}\left(x\mid y\right)a_{k}\left(y\mid z\right)\mathrm{d}y.
\end{align*}
Then, we calculate the right hand side of \eqref{eq:associative_smoothing_app}
we have that
\begin{align*}
a_{i}\otimes a_{jk} & =a'_{ijk},
\end{align*}
where
\begin{align}
a'_{ijk}\left(x\mid z\right) & =\int a_{i}\left(x\mid y\right)a_{jk}\left(y\mid z\right)\mathrm{d}y\nonumber \\
& =\iint a_{i}\left(x\mid y\right)a_{j}\left(y\mid y'\right)a_{k}\left(y'\mid z\right)\mathrm{d}y'\mathrm{d}y.\label{eq:a_ijk_prime_smoothing_app}
\end{align}
We can see that $a_{ijk}$ in \eqref{eq:a_ijk_smoothing_app} is equal
to $a'_{ijk}$ in \eqref{eq:a_ijk_prime_smoothing_app}, which proves
the associative property of the operator in Definition~\ref{def:Operator_smoothing}. 

\subsection{Proof of Theorem~\ref{thm:Smoothing}}

In this appendix, we prove Theorem~\ref{thm:Smoothing}. We first
prove by induction that 
\begin{align}
& a_{k}\otimes\cdots\otimes a_{k+l}\nonumber \\
& =p\left(x_{k}\mid y_{1:k+l},x_{k+l+1}\right), \label{eq:theorem_smo_app1}
\end{align}
for $l<n-k$. Relation \eqref{eq:theorem_smo_app1} holds for $l=0$
by definition of $a_{k}$. Then, assuming that 
\begin{align}
& a_{k}\otimes\cdots\otimes a_{k+l-1}\nonumber \\
& =p\left(x_{k}\mid y_{1:k+l-1},x_{k+l}\right) \label{eq:theorem_filt_app2-1}
\end{align}
holds, we need to prove that \eqref{eq:theorem_filt_app1} holds.

We use $a_{k+l}$ in Theorem~\ref{thm:Smoothing} to calculate
\begin{align*}
& \left[a_{k}\otimes\cdots\otimes a_{k+l-1}\right]\otimes a_{k+l}\\
& =\int p\left(x_{k}\mid y_{1:k+l-1},x_{k+l}\right)p\left(x_{k+l}\mid y_{1:k+l},x_{k+l+1}\right)\mathrm{d}x_{k+l}\\
& =\int p\left(x_{k}\mid y_{1:k+l},x_{k+l},x_{k+l+1}\right)\\
& \quad\times p\left(x_{k+l}\mid y_{1:k+l},x_{k+l+1}\right)\mathrm{d}x_{k+l}\\
& =\int p\left(x_{k},x_{k+l}\mid y_{1:k+l},x_{k+l+1}\right)\mathrm{d}x_{k+l}\\
& =p\left(x_{k}\mid y_{1:k+l},x_{k+l+1}\right).
\end{align*}
This proves \eqref{eq:theorem_smo_app1}. 

If $l=n-k-1$ and $a_{n}$ as in Theorem~\ref{thm:Smoothing}, we
have 
\begin{align*}
& \left[a_{k}\otimes\cdots\otimes a_{n-1}\right]\otimes a_{n}\\
& =\int p\left(x_{k}\mid y_{1:n-1},x_{n}\right)p\left(x_{n}\mid y_{1:n}\right)\mathrm{d}x_{n}\\
& =\int p\left(x_{k}\mid y_{1:n},x_{n}\right)p\left(x_{n}\mid y_{1:n}\right)\mathrm{d}x_{n}\\
& =p\left(x_{k}\mid y_{1:n}\right).
\end{align*}
This result finishes the proof of Theorem~\ref{thm:Smoothing}.

\section{\label{sec:AppendixC}}

In this appendix, we prove Lemma~\ref{prop:Operator_linear_filtering}. We have the following easily verifiable identities:
\begin{align*}
  &\mathrm{N}_I(y ; \eta, J) \mathrm{N}(y ; m,C) \\
  &\propto \mathrm{N}(y ; [J + C^{-1}]^{-1} [\eta + C^{-1} m], [J + C^{-1}]^{-1})  
\end{align*}
and
\begin{align*}
  \mathrm{N}_I(y; \eta, J) \mathrm{N}_I(y; \eta', J') 
  \propto \mathrm{N}_I(y; \eta + \eta', J + J').
\end{align*}
We also have
\begin{align*}
  &\int \mathrm{N}_I(y; \eta, J) \mathrm{N}(y; Az + b, C) \mathrm{d}y \\
  &\propto
  \mathrm{N}_I(z; A^\top [I + J C]^{-1} (\eta - J b) , A^\top [I + J C]^{-1} J A).
\end{align*}
By using Definition~\ref{def:Operator_filtering} for $f_{ij}$ and $g_{ij}$ together with parameterizations in Lemma~\ref{prop:Operator_linear_filtering}, elementary computations lead to \eqref{eq:lin_f} and \eqref{eq:lin_g}.

\section*{Acknowledgment}

The authors would like to thank Academy of Finland for financial support.

\bibliographystyle{IEEEtran}
\bibliography{IEEEabrv,parkf-article}

\end{document}